\documentclass[english]{article}
\usepackage[latin9]{inputenc}
\usepackage[a4paper]{geometry}
\usepackage{textcomp}
\usepackage{graphicx}

\makeatletter
\usepackage{babel}

\makeatother

\usepackage{babel}
\begin{document}
\begin{center}
\textbf{Stabilization and time resolved measurement of the frequency evolution of a modulated diode laser for chirped pulse
generation} 
\par\end{center}

\begin{center}
K. Varga-Umbrich, J. S. Bakos, G. P. Djotyan, P. N. Ignácz, B. Ráczkevi, Zs. Sörlei, J. Szigeti and M. Á. Kedves 
\par\end{center}

\begin{center}
Institute for Particle and Nuclear Physics, Wigner Research Centre for Physics, Hungarian Academy of Sciences; Konkoly-Thege
Miklós út 29-33, H-1121 Budapest, Hungary 
\par\end{center}


\textbf{Abstract}

We have developed experimental methods for the generation of chirped laser pulses of controlled frequency evolution in the
nanosecond pulse length range for coherent atomic interaction studies. The pulses are sliced from the radiation of a cw external
cavity diode laser while its drive current, and consequently its frequency, are sinusoidally modulated. By the proper choice
of the modulation parameters, as well as of the timing of pulse slicing, we can produce a wide variety of frequency sweep
ranges during the pulse. In order to obtain the required frequency chirp, we need to stabilize the center frequency of the
modulated laser and to measure the resulting frequency evolution with appropriate temporal resolution. These tasks have been
solved by creating a beat signal with a reference laser locked to an atomic transition frequency. The beat signal is then
analyzed, as well as its spectral sideband peaks are fed back to the electronics of the frequency stabilization of the modulated
laser. This method is simple and it has the possibility for high speed frequency sweep with narrow bandwidth that is appropriate,
for example, for selective manipulation of atomic states in a magneto-optical trap.


\textbf{Introduction }

The tunability of external cavity diode lasers to hyperfine atomic resonance transitions makes them useful tools for cooling,
trapping and manipulating the atoms \cite{c.e.wieman1991,j.a.pechkis2011,m.j.wright2006,c.affolderbach2005}. These lasers
have narrow bandwidth of about 1 MHz and can be stabilized to an atomic transition. Some experiments for coherently manipulating
the atomic states (for example population transfer, mechanical momentum transfer) often require frequency chirped light pulses
\cite{n.v.vitanov2001,j.s.bakos2007}. The experiments have also shown that the efficiency of the adiabatic population transfer
processes in multilevel atoms strongly depends on the actual range and rate of the frequency sweep during the pulse \cite{j.s.bakos2007}.
Efficient coherent processes can only be induced by laser pulses of carefully adjusted frequency evolution, therefore the
range and rate of the frequency chirp have to be properly stabilized and monitored during the experiment.

In the case of femtosecond laser pulses it is well known that frequency chirping can be achieved simply by inserting dispersive
elements into the optical path \cite{e.b.treacy1969} , however, in the range of nanosecond pulse lengths, these methods
are not feasible and active modulation of the frequency is needed. A straightforward way to produce frequency chirp and light
pulse modulation is to couple an electro optic-phase modulator and an electro-optic amplitude modulator in series with the
laser. The amplitude modulator should be chirp-free (patterned on a so-called X-cut substrate), not to introduce additive
frequency modulation \cite{c.e.rogersiii2007,x.miao2007}. Since large phase modulation is needed for practical purposes,
it is realized by multiple passes through the modulator within an optical loop \cite{c.e.rogersiii2007}. An alternative
method for frequency chirping includes an electro-optic crystal inside the external cavity resonator \cite{k.s.repasky2002,z.cole2004}.

It is possible to create chirped light pulses by applying only one electro-optic amplitude modulator with Z-cut crystals
which have the property of producing frequency sweep synchronously, proportional to the amplitude modulation \cite{j.s.bakos2009}.
However, in this case the frequency excursion is bond with the amplitude variation, so they cannot be varied independently.
Similar parasitic phase modulation occurs in the phase shifting interferometry, where the nonlinear response of a piezoelectric
transducer (PZT) leads to chirped signal detection on the CCD \cite{r.langoju2007}.

In the cases mentioned above, i.e. when the laser diode is operated at a constant current, the laser frequency can be stabilized
by one of the typical spectroscopic methods \cite{c.e.wieman1991,g.galzerano2006,s.wu2007}, where some part of the laser
light passes through an absorption cell. The narrow absorption signal is used for side-locking or peak-locking by an electronic
controller of the laser frequency. Pound-Drever-Hall technique is an improved method for frequency stabilization \cite{e.d.black2001},
where the frequency is measured with a Fabry-Perot cavity with sharp transmission (or reflection) lines and this result is
fed back to the laser PZT to maintain the frequency. A robust frequency stabilization technique uses the Zeeman-shift in
atomic Doppler broadened absorption signals which offers large recapture range and rarely loses lock even in an extremely
noisy environment \cite{k.l.corwin1998}.

Another possibility to produce frequency modulated laser pulses on the nanosecond time scale is the modulation of the frequency
of the laser diode combined with an external device, e.g. an amplitude modulator, for pulse shaping. The injection current
modulation is fast enough to achieve tens of MHz repetition rate of the laser frequency modulation, but the optical frequency
change is limited by modes hops. To reach the required frequency range one has to vary the external resonator length with
a PZT \cite{l.ricci1995}. Moreover the optical feedback from the grating narrows the linewidth to below 1 MHz and the PZT
of the grating is controlled by an electronic system for frequency tuning and locking. However, the PZT is not an appropriate
device for operating on the nanosecond time scale because of mechanical reasons, that is why the current modulation seems
to be the only approach to creating the desired frequency modulated signal in the nanosecond range of duration.

Current modulation causes both intensity and frequency modulation. Intensity modulation can be alleviated by injection locking
of a separate laser with the modulated light \cite{k.sasaki2006,k.szymaniec1997}, where the slave laser reproduces the frequency
modulation only.

The stabilization of the frequency evolution of a modulated laser requires more complicated solutions. The carrier frequency
has to be locked to an appropriate reference frequency in this case. There are several papers on the mean frequency stabilization
of current modulated semiconductor lasers using the modulation sidebands of a Fabry-Perot interferometer as frequency references
for the feedback loop \cite{c.j.nielsen1983,h.tsuchida1987,h.s.lee1990}. In those setups, however, the speed of the frequency
modulation is limited by the ringing of the Fabry-Perot interferometer for light pulses shorter than the transition time
of the interferometer \cite{l.ricci1995}. This ringing is caused by the oscillations of the field inside the interferometer,
and results in non-monotonic frequency change of the transmitted light pulse \cite{m.s.fee1992}.

In the present paper we describe an experimental system for generating frequency chirped nanosecond laser pulses of controlled
frequency evolution. Frequency modulation is achieved by sinusoidal modulation of the diode laser current, and the pulses
are sliced by an X-cut substrate amplitude modulator for chirp-free operation. For stabilization the spectral lines of the
beat signal between the modulated and a stabilized constant frequency laser are used, in order to overcome the limitation
of ringing effects in a Fabry-Perot interferometer.

\textbf{Frequency lock of a modulated laser}

In our measurements two single mode diode lasers are used for the realization of the stabilized chirped-pulse system. The
experimental arrangement is illustrated in Fig.1.

\begin{center}
\includegraphics[scale=0.55]{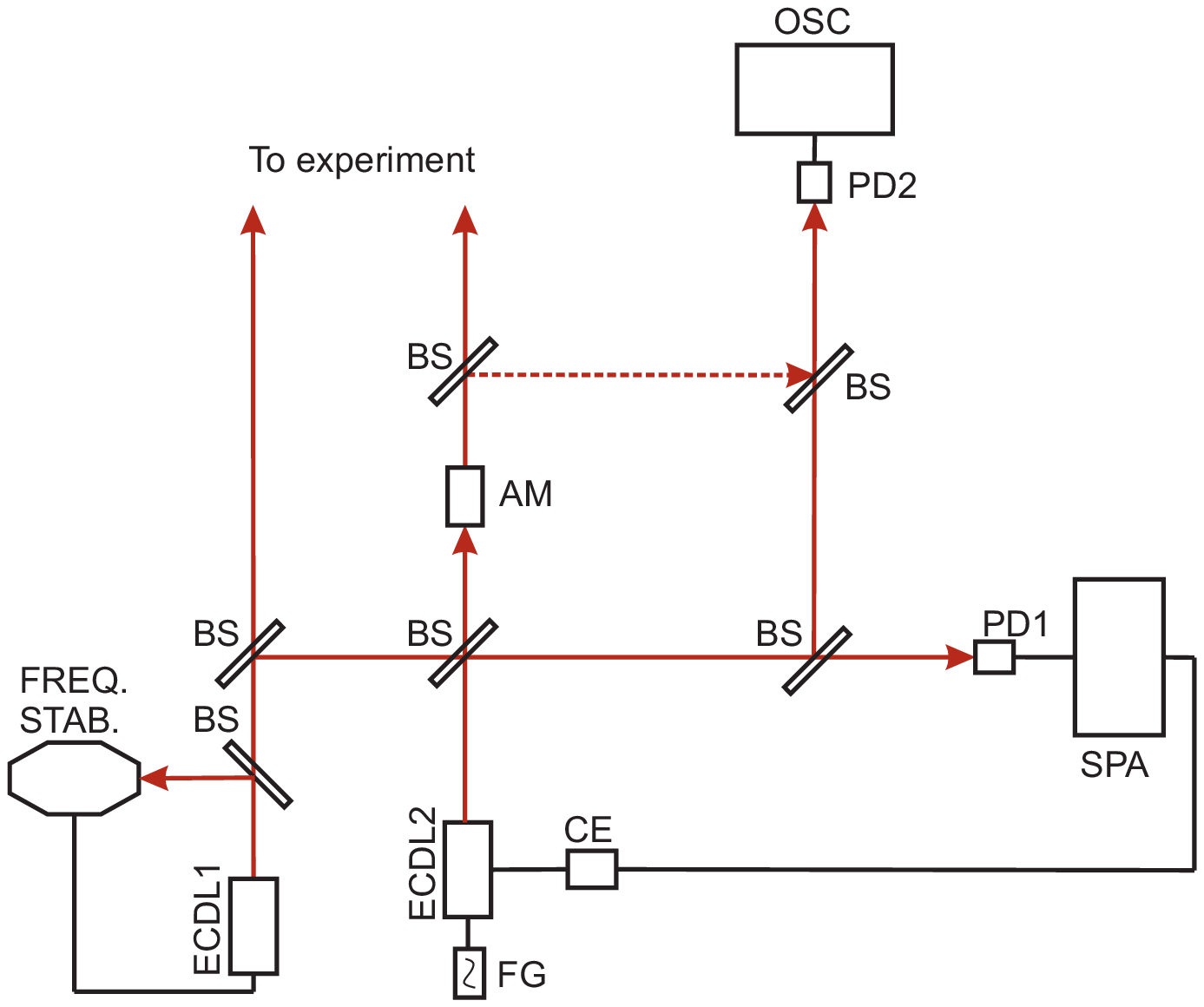}
\par\end{center}

\textbf{\small{}{}Fig.1.}{\small{}{} Schematic experimental setup for the stabilization and measurement of the frequency
of a modulated external cavity diode laser. ECDL1: reference laser stabilized by saturation spectroscopy (FREQ.STAB.); ECDL2:
modulated laser locked to the reference laser frequency; FG: function generator; SPA: spectrum analizer; CE: control electronics
(including a PID regulator) for stabilizing the center frequency of ECDL2; AM: amplitude modulator for slicing the pulses
from the cw radiation of ECDL2; OSC: fast digital oscilloscope; PD1, PD2: fast photodiodes; BS: beamsplitters.}{\small \par}

The reference light source (ECDL1) is an EOSI 2001 (bandwidth: 100 kHz) external cavity diode laser with a Littmann-Metcalf
resonator stabilized by saturation absorption spectroscopy to the F=3-\textgreater{}F' hyperfine transitions of the D2
line of $^{85}$Rb at 780 nm wavelength \cite{c.affolderbach2005}. The sub\textendash Doppler absorption signal is provided
by a rubidium vapour cell and it is held on the top-of-fringe by a lock-in amplifier. The frequency modulated laser (ECDL2)
is a Toptica DL 100 external cavity diode laser with a Littrow resonator, and the feeding current is modulated with a function
generator (FG) through a Bias-T coupler. The lasers are mounted on metal heat sinks with Peltier-element cooling. It was
necessary to operate the lasers at a temperature between 18 \textdegree C and 19 \textdegree C to obtain single mode operation
while the room temperature in the laboratory was set to be 21 \textdegree C during the experiments.

Beams from the lasers were joined at a beamsplitter (BS) and the beat wavefront with an optical power of 0.1-1 mW was split
in two and observed by fast photo detectors (PD1: Menlo Systems APD110: 1-800 MHz, and PD2: New Focus 1591: 4.5 GHz bandwidth).
One of the signals was then registered by a Tektronix DPO7104 digital oscilloscope (1.0 GHz bandwidth; OSC), and the other
one was analyzed by a Takeda Riken 110 spectrum analyzer (120 MHz bandwidth; TS). The Fourier spectrum of the interference
signal (Fig.2) is produced in this spectrum analyzer and the narrow spectral lines can be used for frequency stabilization.

The main output beam of the modulated laser was then passed through an integrated Lithium-Niobate amplitude modulator (Photline
NIR-MX800) which sliced the appropriately shaped pulses from the continuous wave radiation. The resulting length, shape and
timing parameters of the pulses were regularly measured by optionally coupling part of the beam into the PD2 photodetector.
This beam, after amplification, was sent to the experiment for inducing the coherent adiabatic transitions in Rubidium atoms.

\begin{center}
\includegraphics[scale=0.55]{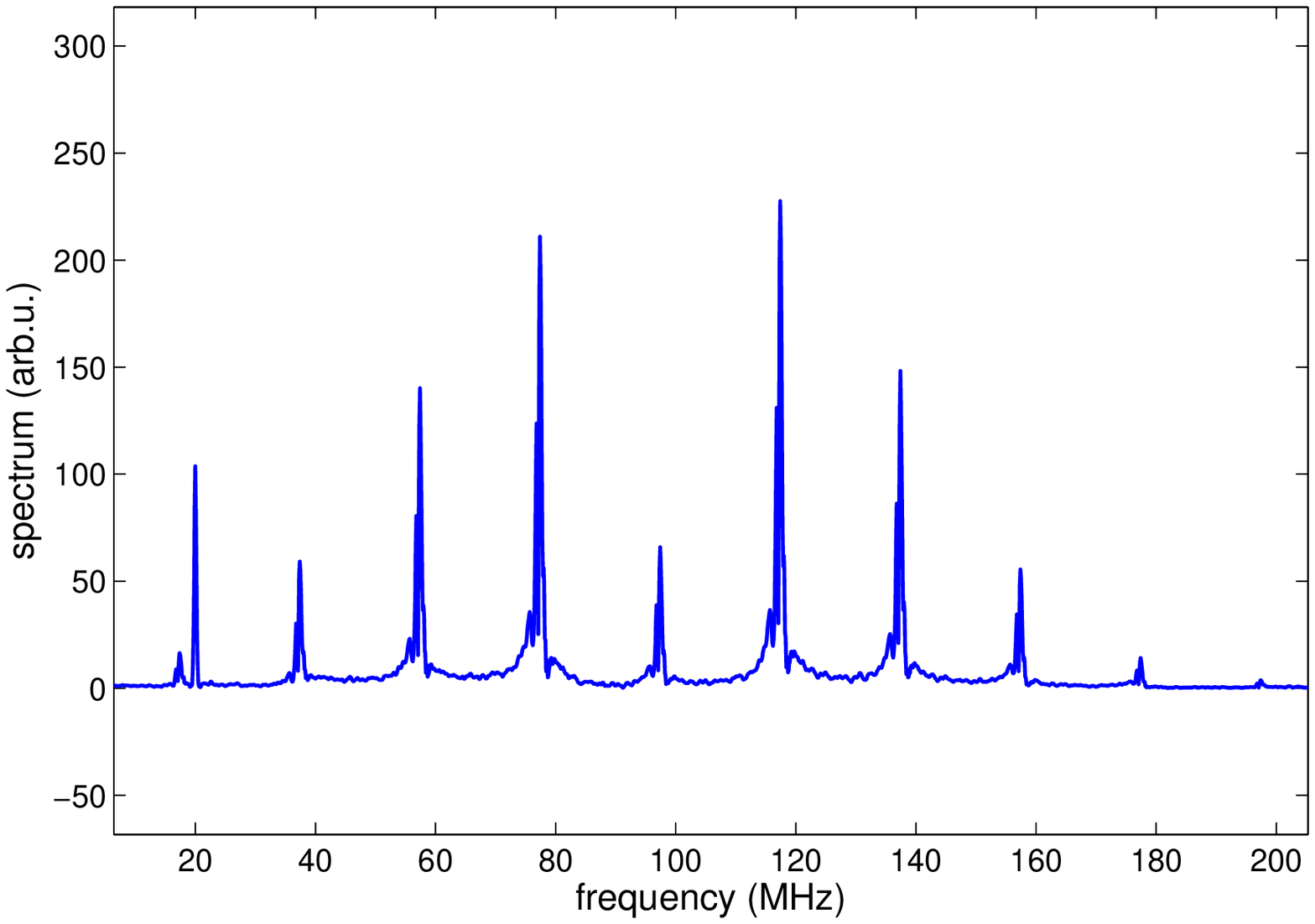}
\par\end{center}

\begin{center}
\textbf{\small{}{}Fig.2. }{\small{}{} Positive frequency part of the absolute value of the Fourier spectrum of the interference
signal between the reference and modulated lasers. The modulation frequency was 20 MHz.} 
\par\end{center}

To avoid the overlap between the interference spectrum lines it is necessary to apply a beat frequency larger than the width
of the lines, that is, in the MHz range for external cavity lasers. We used the spectrum analyzer to filter out a selected
frequency peak and put out the signal for the stabilizing electronics of the laser. Since the measurements were performed
near the rubidium atomic resonance where the reference laser's frequency is locked, there is always a peak in the beat spectrum
which falls into the spectrum analyzer bandwidth, so we can choose it for stabilization.

By switching off the frequency sweep of the spectrum analyzer it is used as a bandpass filter which gives a positive or negative
voltage on its `Y` output according to the input beat signal frequency being below or above the adjusted bandpass filter
middle point. The wavelength tuning can be achieved by moving the bandpass window through the spectrum and locking different
sidebands. The narrow and noisy peaks in the Fourier spectrum are smoothed by the built-in electronic filter in order to
make them appropriate for stabilization.

While top-of-fringe locking would be possible to lock the signal level with a lock-in amplifier, we applied side-of-fringe
locking for stabilization, because the laser has a PID-module (proportional, integral, differential-filter). This electronic
module produces a feedback current proportional to the difference of the `Y` output of the spectrum analyzer and a reference
level set previously. The feedback loop is able to correct the frequency drift and low frequency noise due to thermal and
mechanical instabilities. The output signal of this stabilization electronics is then coupled to the frequency control inputs
of the diode laser, i.e. the piezo voltage which determines the angle of the grating of the resonator, as well as the feed
current of the laser diode.

In principle, any peak could be selected, so this method can also be used for stabilization as far as some hundreds of MHz
from an atomic resonance if it is necessary. Moreover, the selected peak can be shifted by electronic mixing inside the spectrum
analyzer that means fine wavelength tunability within the 100 MHz range. This versatility is useful for example in capturing
atoms in optical traps, where the frequency blueshift of the counterpropagating radiation in the frame of the escaping atoms
can be easily compensated by setting the laser frequency below the atomic resonance frequency.

The operation of the stabilization scheme was first inspected without modulating the locked laser. Long beat signals consisting
of 50 million points were recorded and the Fourier spectrum of the signals were calculated for sections of different lengths.
The bandwidth of the signals of different lengths was determined by fitting a Lorentzian function to the spectra. (It should
be noted that for long sections of the unstabilized signals the spectra deviated from a Lorentzian shape due to random drifts;
the increase of the calculated bandwidths, however, indicate the significant changes in the difference frequency.) The dependence
of the obtained bandwith values on the length of the signal sections demonstrates the timescale of the stabilzation procedure.
The results with and without locking can be seen in Fig.3. (The frequency of the reference laser was not stabilized in this
case.)

\begin{center}
\includegraphics[scale=0.55]{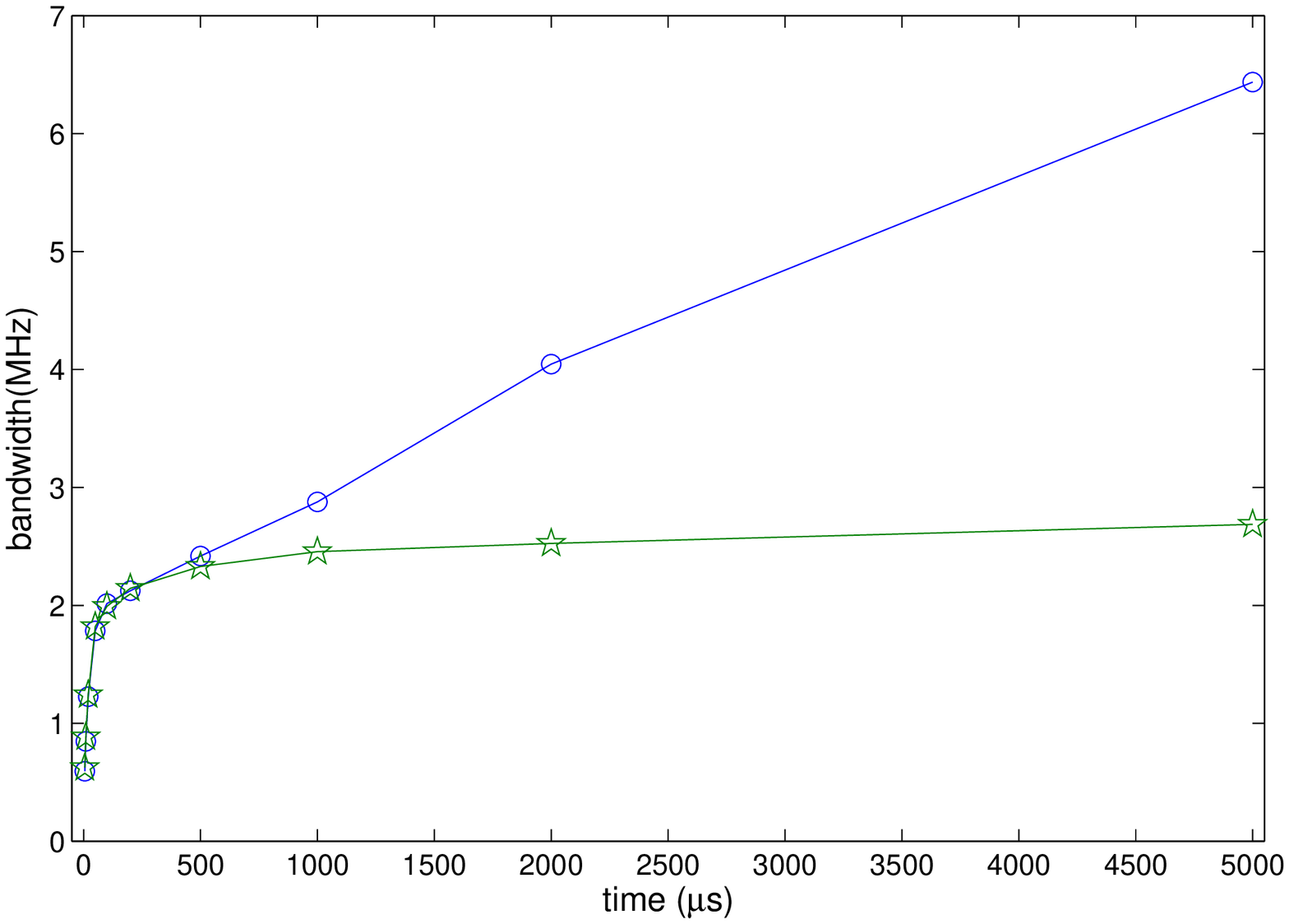}
\par\end{center}

\begin{center}
\textbf{\small{}{}Fig.3.}{\small{} {} Spectral bandwidth values of the beat frequency obtained for the two lasers without
modulation. The frequency spread calculated for time intervals of different lengths is displayed versus the length of the
period, for the stabilized (stars) and unstabilized (circles) cases. }
\par\end{center}{\small \par}

{\small{}The comparison reveals that the locking is efficient since the variation of the frequency of the stabilized laser
does not increase with time after about 0.5 ms, in contrast to the unlocked case. This result also shows the time scale of
the feedback which is about 0.5 ms where the two curves diverge and approximately corresponds to the frequency limit of the
piezo actuator of the grating of the laser resonator.}{\small \par}

\textbf{\small{}Measurement of the modulated frequency evolution}{\small \par}

{\small{}In order to check the efficiency of the stabilization method, as well as to determine the frequency evolution of
the laser radiation during the nanosecond pulses, we performed time resolved measurement of the modulated laser's frequency.
Since the feed current of the laser was modulated sinusoidally with a frequency $f_{mod}$, we expected a sinusoidal variation
of the laser frequency too (provided that the modulation amplitude was small enough to avoid mode-hops): $\nu_{m}=\nu_{m0}-\nu_{mod}\cdot sin\left(2\pi\cdot f_{mod}\cdot t+\varphi_{m}\right)$.
The parameters of the function describing the frequency variation were determined by least squares fitting of the appropriate
mathematical function to the interference signal recorded by the digital oscilloscope OSC.\cite{r.t.white2004}}{\small \par}

{\small{}Since the constant and the chirped frequency laser fields are linearly polarized plane waves, we get the following
expression for the detected interference intensity:}{\small \par}

\begin{center}
{\small{}$I=I_{r}+I_{m}+2\sqrt{I_{r}I_{m}}\cdot\delta\cdot cos\left(\frac{\nu_{mod}}{f_{mod}}cos\left(2\pi\cdot f_{mod}\cdot t+\varphi_{m}\right)+2\pi\cdot\Delta\nu\cdot t+\Delta\varphi\right)$ }
\par\end{center}{\small \par}

{\small{}where the first term Ir is the intensity of the reference laser beam alone, $I_{m}=I_{m0}+I_{mod}\cdot cos\left(2\pi\cdot f_{mod}\cdot t+\varphi_{I}\right)$
is the time dependent intensity of the modulated laser beam, and the $\delta$ modulation index is the interference `efficiency`
that takes into account the imperfect overlap of the two beams. $\Delta\nu=\nu_{m0}-\nu_{r}$ represents the difference between
the carrier frequency (in other words, the constant component or mid-value of the frequency) of the modulated laser and the
reference laser's frequency. This is the physical quantity we would like to stabilize in our experiments.}{\small \par}

{\small{}In order to fully characterize the frequency evolution of the modulated laser radiation, we also have to determine
the amplitude: $\nu_{mod}$ , and phase: $\varphi_{m}$ of the frequency modulation. To complete the fitting procedure, all
the remaining parameters also have to be determined. The intensity signals of the two lasers $(I_{r},I_{m})$ were measured
separately too, and the interference efficiency value $(\delta)$ was calculated from the beat signal detected without modulation
of the ECDL2. $f_{mod}$ is known in advance and can also be obtained precisely from the oscilloscope signal of the driving
current. The fitting procedure is used to calculate $\Delta\nu,\nu_{mod},\varphi_{m}$ and $\Delta\varphi$ , the latter
being the phase difference between the two optical fields. The result of the calculation together with the measured interference
intensity is demonstrated in Fig.4 showing an actual beat signal together with the fitted curve.}{\small \par}

\begin{center}
\includegraphics[scale=0.55]{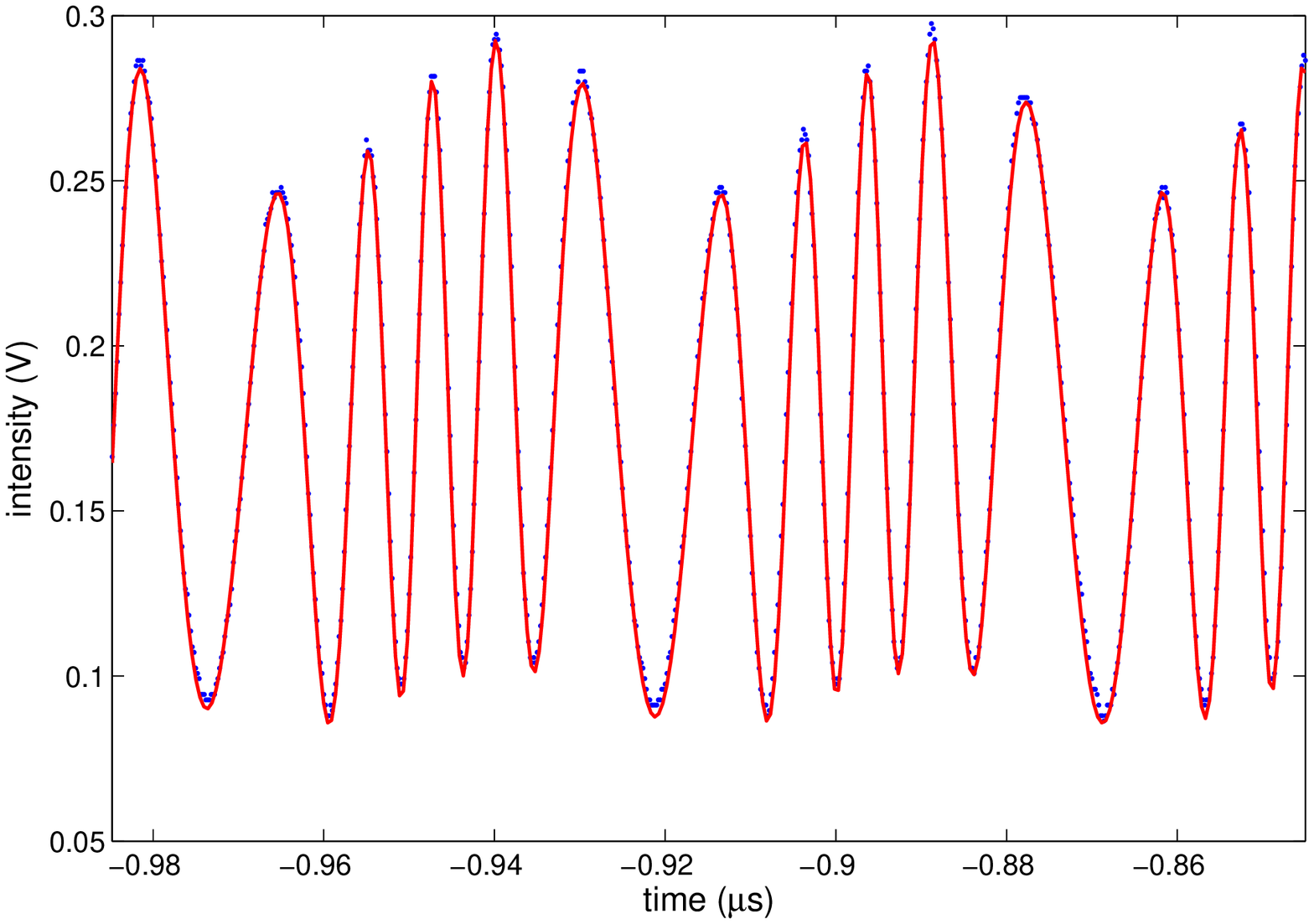}
\par\end{center}

\begin{center}
\textbf{\small{}{}Fig.4.}{\small{} {} Short section of the measured beat signal (dots) with the fitted interference function
(solid line). }
\par\end{center}{\small \par}

{\small{}The evolution of the modulated laser's frequency during the chirped pulse is of primary importance from the point
of view of the coherent atomic excitation in the experiment. Therefore, the phase of the modulated laser's frequency variation
had to be carefully synchronized with the timing of the pulse slicing. The output pulse train from the amplitude modulator
was regularly measured by coupling this beam into the fast photodetector PD2, and the phases of the signals were synchronized
by the driving modulation of the function generator. The frequency chirp during the modulated laser's pulses determined by
this measurement procedure is illustrated in Fig.5, where the pulse shape can be seen together with the reconstructed frequency
evolution.}{\small \par}

\begin{center}
\includegraphics[scale=0.55]{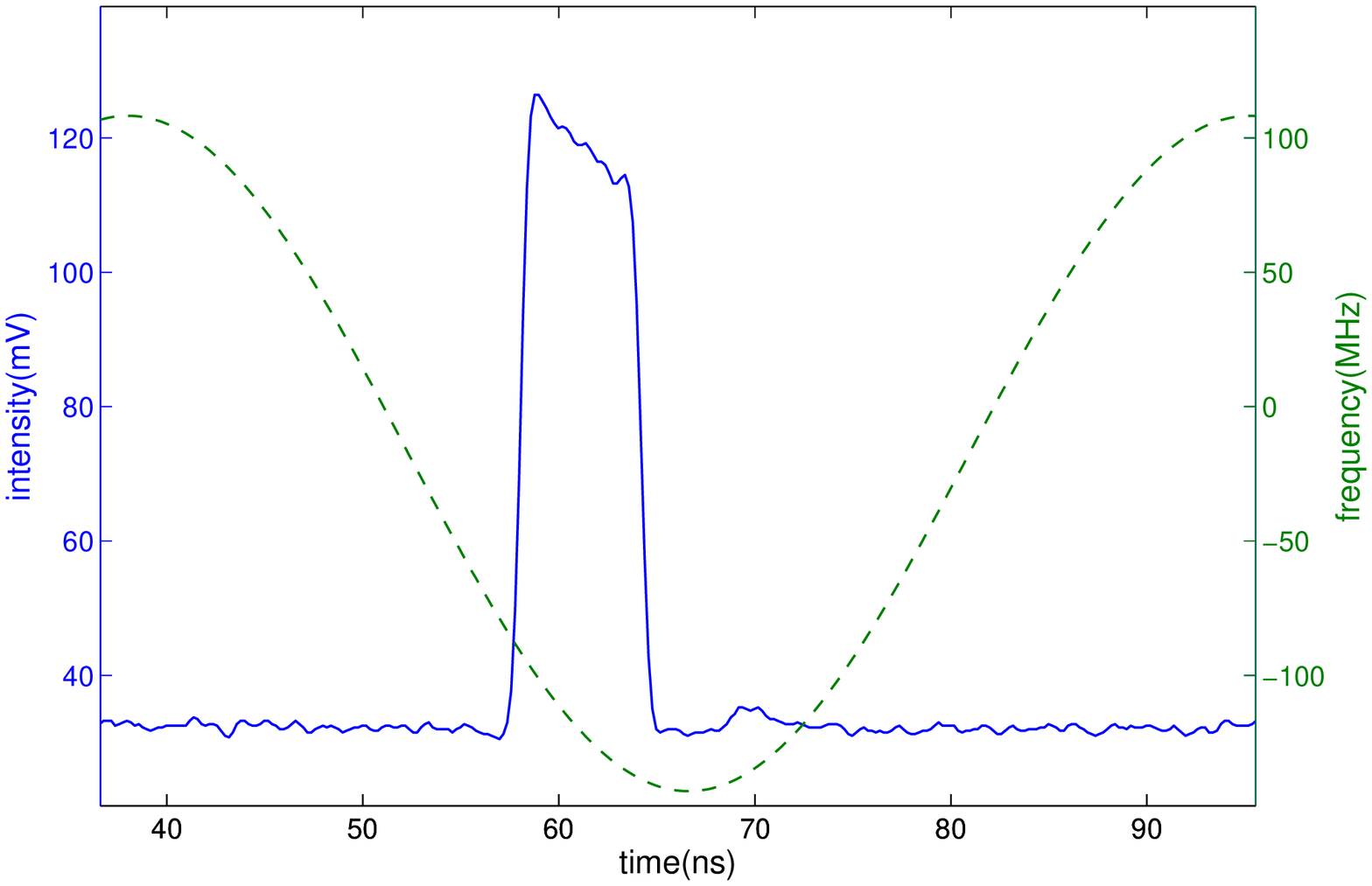}
\par\end{center}

\begin{center}
\textbf{\small{}{}Fig.5.}{\small{} {} Laser pulse shape (continous line, left axis) together with the frequency evolution
(dashed line, right axes) obtained by the interference measurement. }
\par\end{center}{\small \par}

{\small{}In order to validate the efficiency of the developed stabilization technique, numerous measurement series have been
completed both with and without switching on the frequency lock.}{\small \par}

\begin{center}
\includegraphics[scale=0.55]{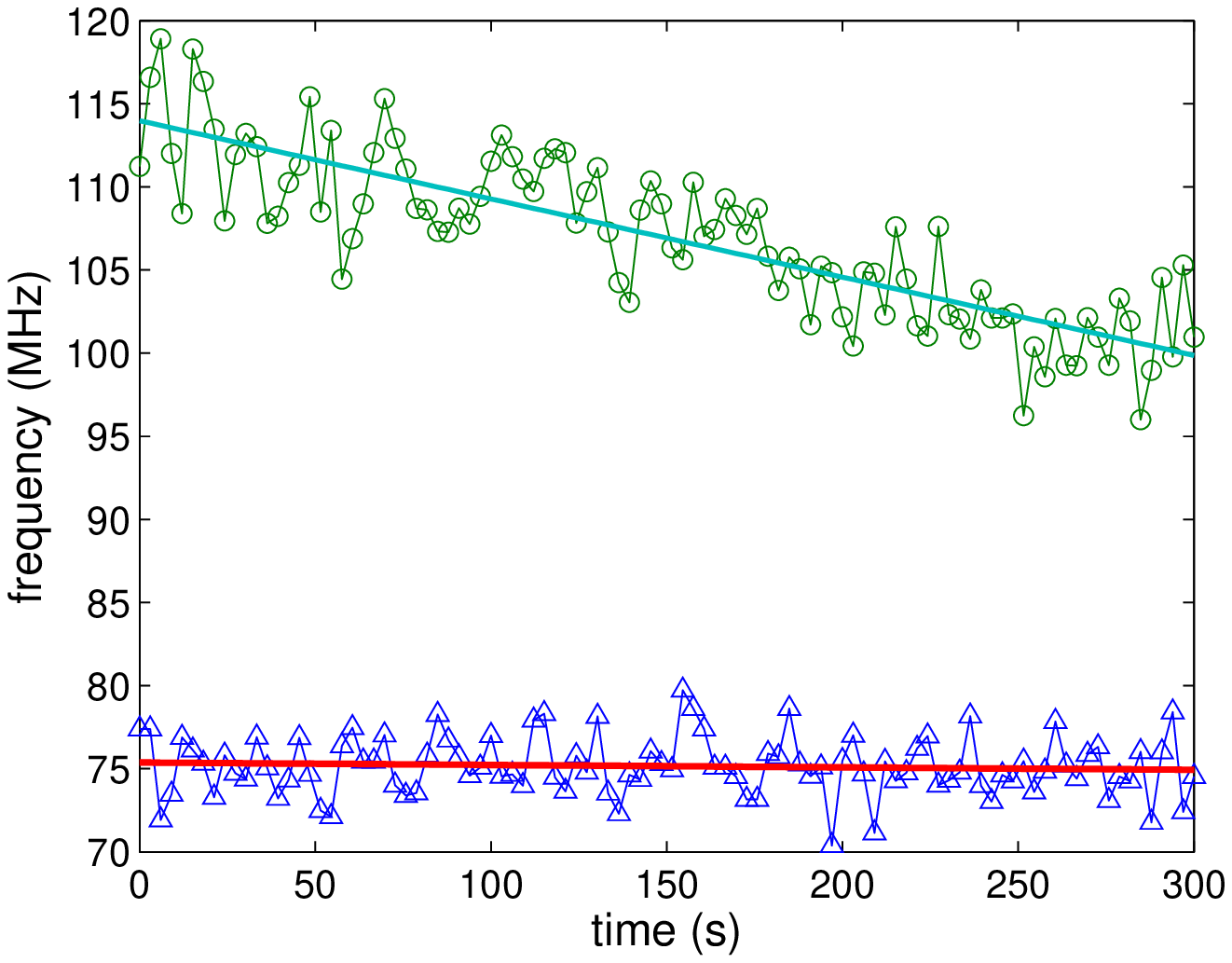}
\par\end{center}

\begin{center}
\textbf{\small{}{}Fig.6.}{\small{} {}Results of the stabilization procedure: variation of the carrier frequency values
of the modulated laser with time, when the locking was on (triangles) and off (circles). The continuous lines show the linear
fits indicating the frequency drift. }
\par\end{center}{\small \par}

{\small{}The interference signals were recorded every few seconds for several minutes, i.e. some hundreds of measurement
points were taken. The carrier frequency values of the modulated laser were then reconstructed by the fitting procedure described
above. The results of these experiments are illustrated in Fig.6 where the mid-frequency values obtained in a series of 100
points with and without stabilization can be seen. The Figure shows that the frequency variation is significantly less in
the stabilized case compared with the case of no stabilization. As it can be seen in Fig.6, the frequency is drifting for
the free-running laser with a rate equal to 47 kHz/s, and the total frequency drift was 14.1 MHz in 300 seconds, which is
not acceptable in precise applications. At the same time, the frequency of the locked laser drifted 440 kHz in 300 s, which
corresponds to 1.4 kHz/s drift rate. Similar results have been obtained in repeated measurement series too: the frequency
drift without stabilization was several times 10 kHz/s (50 kHz/s or even higher), while only a few kHz/s in the stabilized
case. The remaining drift can most probably be attributed to the controlling scheme of the feedback which locks to a signal
level; this drift can be eliminated by applying a more complicated arrangement to realize top-of-fringe locking using lock-in
techniques. In order to estimate the frequency scatter we have also calculated the Allan deviation of the resulted values
of the series. The absolute (i.e. not normalized to the signal frequency,) Allan deviation without locking was 2.37 MHz and
for the locked laser it was 1.66 MHz, which also indicates a decrease in frequency variation. This result depends mainly
on the controlling electronic system used for stabilization and may be improved substantially by further developing of the
electronic system.}{\small \par}

\textbf{\small{}Conclusion}{\small \par}

{\small{}We have realized a simple, versatile and robust method of frequency locking of an external cavity diode laser to
produce frequency chirped narrowband light pulses in the nanosecond time scale. In our setup the speed of the frequency chirp
is not limited by the ringing of the Fabry-Perot resonator. To overcome this limitation, the frequency modulated laser light
is mixed with another frequency stabilized laser light and the lines of the beat spectrum are used for stabilization. The
stabilized frequency can be shifted in the hundred MHz range by tuning the band-pass filter to ensure the appropriate frequency
for atom-optic experiments. The frequency evolution of the stabilized laser is measured by evaluating the beat signal with
the reference laser radiation. The long term stability of the mid-frequency of the modulated laser has been demonstrated
experimentally.}{\small \par}

\textbf{\small{}Acknowledgement}{\small \par}

{\small{}This work was funded by the Research Fund of the Hungarian Academy of the Sciences (OTKA) under contracts K 68240,
ELI-09-1-2010-0010 grant, and the grant for the Centre of Excellence of the Hungarian Academy of Sciences.}{\small \par}

{\small{}\bibliographystyle{unsrt}
\bibliography{Carrier_frequency_locking_2015_0419}
 }
\end{document}